\documentstyle[12pt]{article}
\columnwidth\textwidth
\newcommand{\goes}[1]{\stackrel{#1}{\longrightarrow}}

\newcommand{\Longgoes}[1]{\stackrel{#1}{-\!\!\!-\!\!\!\longrightarrow}}

\newcommand{\nsymbol}{{\mbox{$I \! \! N$}}}
\begin{document}

\begin{center}
{\large\bf Decidability Issues for Petri Nets}\\[4mm]
{\bf -- a survey\footnote{A former version of this paper
appeared in the Concurrency Column of the EATCS Bulletin 52.}}\\[10mm]
\begin{tabular}{cc}
Javier Esparza$^*$& Mogens Nielsen\\
Institut f\"ur Informatik& BRICS$^{**}$\\ Technische Universit\"at
M\"unchen
 &
Computer Science
Department \\
Arcisstr. 21 & Aarhus University\\
D-80290 M\"unchen & Ny Munkegade, Bldg. 540\\
Germany &  DK-8000 Aarhus C\\
 & Denmark
\end{tabular}
\end{center}

\begin{abstract}
We survey 25 years of research on decidability issues for Petri nets. We collect 
results on the decidability of important properties, equivalence notions, 
and temporal logics. 
\end{abstract}

\section{Introduction}
Petri nets are one of the most popular formal models for the
representation and analysis of parallel processes. They are due to
C.A.\ Petri, who introduced them in his doctoral dissertation in
 1962.
Some years later, and independently from Petri's work, Karp and
Miller introduced vector addition systems \cite{KarpMiller69}, a
simple mathematical structure which they used to analyse the
properties of `parallel program schemata', a model for parallel
computation. In their seminal paper on parallel program schemata,
Karp and Miller studied some decidability issues for vector
addition systems, and the topic continued to be investigated by other
researchers. When Petri's ideas reached the States around 1970,
 it was
observed that Petri nets and vector addition systems were
mathematically equivalent, even though their underlying philosophical
ideas were rather different (a computational approach to the
physical world in Petri's view, a formal model for concurrent
programming in Karp and Miller's). This gave more momentum to the
research on decidability questions for Petri nets, which has
since continued at a steady pace.

In the following we have collected some highlights of decidability
 issues
for Petri nets from
the 70's, 80's and 90's. As you will see, they form a nice mixture
 of old
celebrated
breakthroughs, and a recent burst of exciting new developments.

We have decided to group our selected results in three sections,
 covering
respectively the
decidability of specific {\em properties}, various (behavioural)
 {\em
equivalences}, and finally
the model checking problem for {\em temporal logics}.

\rule{50mm}{.01in}\\
{\footnotesize $^*$ This paper was written while this author was at the University of
Edinburgh.}\\ {\footnotesize $^{**}$ Basic Research In Computer Science, Centre
of the Danish National Research Foundation.}

\pagebreak

It should be noted that we have selected our highlights also aiming
 at some
coherence in our
presentation. In other words, we do not claim to cover all important
contributions on
decidability for Petri nets, but still our selection covers a pretty
comprehensive part of
existing results, also compared to other similar surveys, e.g.\
\cite{Jantzen86}. We have not included results on
extensions of the Petri net model. In particular, for decidability results
on timed Petri nets we refer the reader to
\cite{JonesLandweberLien77,ValeroFrutosCuartero91,ValeroFrutosCuartero93}.

\section{Basic definitions}

We give, in a somewhat informal way, the basic definitions on Petri
 nets
that we need in order to state the results of this overview.

A {\em net} $N$ is a triple $(S, T, F)$, where
$S$ and $T$ are two disjoint, finite sets,
and $F$ is a relation on $S \cup T$ such that $F \cap (S \times S)
 = F \cap (T
\times T) = \emptyset$. The elements of $S$ and $T$ are called
 {\em places} and
{\em transitions}, respectively, and the elements of $F$ are called
 {\em
arcs}. A {\em marking} of a net $N = (S,T,F)$ is a mapping  $M
 \colon S \to
\nsymbol$. A marking $M$ {\em enables} a transition $t$ if it marks
 all its
input places. If $t$ is enabled at $M$, then it can {\em occur},
 and its
occurrence  leads to the  successor marking $M'$,  which is defined
 for every
place $s$ as follows: a token is removed from each input place of
 $t$ and a
token is added to each output place of $t$ (if a place is both input
 and
output place of a transition, then its number of tokens does not
 change).
This is denoted by $M \goes{t} M'$.

A {\em Petri net} is a pair $(N, M_0)$, where $N$ is a net and $M_0$
 a marking
of $N$, called {\em initial marking}. A sequence $M_0 \goes{t_1}
 M_1 \goes{t_2}
\cdots \goes{t_n} M_n$ is a {\em finite
occurrence sequence} leading from $M$ to $M_n$
and we write $M_0 \Longgoes{t_1 \ldots t_n} M_n$.
A sequence  $M_0 \goes{t_1} M_1 \goes{t_2} \cdots$ is an {\em infinite
occurrence sequence}.
An occurrence sequence is {\em maximal} if it is infinite, or it leads
to a
marking which does not enable any transition. A marking $M$ of $N$
 is {\em
reachable} if $M_0 \goes{\sigma} M$ for some sequence $\sigma$.
The {\em reachability graph} of a Petri net is a labelled graph
whose nodes are the reachable markings; given two reachable markings
 $M$,
$M'$, the reachability graph contains an edge from $M$ to $M'$ labelled
 by a
transition $t$ if and only if $M \goes{t} M'$.

A {\em labelled} net is a quadruple $(S, T, F, \ell)$, where $(S,
 T, F)$ is a
net and $\ell$ is a labeling function which assigns a letter of some
 alphabet
to each transition. This function need not be injective. The reachability
 graph
of a labelled net is defined like that of unlabelled nets; the only
difference
is that if $M \goes{t} M'$ then the corresponding edge from $M$ to
$M'$ is
labelled by $\ell(t)$.

Sometimes we refer to
the `normal' Petri nets as {\em unlabelled} Petri nets. Unlabelled Petri
nets can also be seen as labelled Petri nets in which the labelling
function is injective.

Given a Petri net $(N,M_0)$ and a marking $M^f$ of $N$ (called {\em
 final
marking}), we define the {\em language} of $(N, M_0)$ with respect
 to $M^f$ as
\begin{eqnarray*}
L(N, M_0, M^f) &=  & \{\sigma \mid M_0 \goes{\sigma} M^f\}
\end{eqnarray*}
\noindent and the {\em trace set} of $(N, M_0)$ as
\begin{eqnarray*}
T(N, M_0) &=  & \{\sigma \mid M_0 \goes{\sigma} M \mbox{ for some
 marking $M$}
\} \end{eqnarray*} (sometimes the terms `language' and `terminal language'
are used instead of `trace set'
 and `language'). Please note that the term  `trace' is also used
in the theory of Petri nets for elements of a free partially commutative
monoid, following an idea originally due to Mazurkiewicz \cite{Maz88}. In
order to focus our presentation, we have chosen not to include
decidability results on `Mazurkiewicz traces'.

Given a labelled Petri net $(N, M_0)$, where $N= (S,T,F,\ell)$,
 and a marking
$M^f$ of $N$, the
 {\em language} of $(N, M_0)$ with respect to $M^f$ as
\begin{eqnarray*}
L(N, M_0, M^f) &=  & \{\ell(\sigma) \mid M_0 \goes{\sigma} M^f \}
\end{eqnarray*}
 \noindent and the  {\em trace set} of $(N, M_0)$ is defined as
\begin{eqnarray*}
T(N, M_0) &=  & \{\ell(\sigma) \mid M_0 \goes{\sigma} M \mbox{ for
 some marking
$M$}\} \end{eqnarray*}
We now define those classes of nets that are mentioned several times
 along the
survey. Some others, which appear only once, are defined on the fly (or
a reference is given).

A Petri net $(N, M_0)$ is:
\begin{itemize}
\item {\em persistent} if for any two different transitions $t_1$,
 $t_2$ of
$N$ and any reachable marking $M$, if $t_1$ and $t_2$ are enabled at
$M$, then the occurrence of one cannot disable the other.
\item {\em conflict-free} if, for every place $s$ with more than
 one
output transition, every output transition of $s$ is also one of
 its
input transitions. All conflict-free nets are persistent; in fact,
 $(N,
M_0)$ is conflict-free if and only if $(N, M)$ is persistent for
 every
marking $M$ of $N$. For most purposes, this class is equivalent to
the class of nets in which each place has at most one output transition.
\item {\em sinkless} if any
cycle of $N$ which is marked at $M_0$ (meaning that $M_0(s) >0$ for
 some
place of the cycle) remains marked at every reachable marking; i.e.,
 cycles
cannot be emptied of tokens by the occurrence of transitions.
\item {\em normal} if, for any cycle of the net, every input transition
 of
some place of the cycle is also an output transition of some place
 of the
cycle. All normal nets are sinkless; in fact, normal is to sinkless
what conflict-free is to persistent: $(N, M_0)$ is normal if and
 only if
$(N, M)$ is sinkless for every marking $M$ of $N$.
\item {\em single-path} if it has a unique maximal occurrence sequence.
\item a {\em BPP-net} if every transition has exactly one input place.
 BPP
stands for Basic Parallel Process. This is a class of CCS processes
defined by Christensen \cite{Christensen93} (see also the Concurrency
 column
of the EATCS Bulletin 51). BPPs can be given a net semantics in terms
 of BPP-nets. BPP-nets are also computationally equivalent to
commutative context-free grammars, as defined by Huynh in
\cite{Huynh83}.

\item {\em free-choice} if, whenever
an arc connects a place $s$ to a transition $t$, either $t$ is the
unique output transition of $s$, or $s$ is the unique input place
 of
$t$.
\item {\em 1-safe} if, for every place $s$ and every reachable marking
 $M$,
$M(s) \leq 1$; i.e., no reachable marking ever puts more than one
 token in any
place.
\item {\em symmetric} if for every transition
$t$ there is a reverse transition $t'$ whose occurrence `undoes' the
effect of the occurrence of $t$, i.e., the input places of $t$ are the
output places of $t'$, and vice versa.
\item {\em cyclic} if $M_0$ can be reached from any reachable marking;
i.e., it is always possible to return to the initial
marking.\footnote{Symmetric and cyclic nets have sometimes been called
{\em reversible}.}
\end{itemize}

\section{Properties}
\label{properties}
In spite of the rather large expressive power of Petri nets, we shall
 see
in this section that most of the usual properties of interest for
 verification
purposes are decidable. On the other hand, we shall also see that
 they
tend to have very large complexities. In fact, Petri nets are an important
source of natural non-primitive recursive problems!

So far, all decidability proofs in the net literature are
carried out by reduction to the boundedness or the reachability
problem: these are the only two with a direct decidability proof,
 and we are
thus obliged to begin the section with them.

\paragraph{Boundedness}

A Petri net is {\em bounded} if its set of
reachable markings is finite. Karp and Miller proved in
\cite{KarpMiller69} that boundedness is decidable. This result follows
from the following characterization of the unbounded Petri nets,
not difficult to prove. A
Petri net is unbounded if and only if there exists a reachable
marking $M$ and a sequence of transitions $\sigma$ such that $M
\goes{\sigma} M + L$, where $L$ is some non-zero marking, and the sum of
markings is defined place-wise. The sequence $\sigma$ is a sort of `token
generator' which, starting from a marking $M$, leads to a bigger one $M +
L$.

Karp and Miller showed how to detect `token generators' by constructing
what was later called the {\em coverability tree}.
Their algorithm turns out to be surprisingly inefficient: token
generators may have non-primitive recursive length in the size of
the Petri net, which implies that the construction of the coverability
tree requires non-primitive recursive space! Rackoff gave a better
algorithm in \cite{Rackoff78}. He showed that there always
exists one token generator of `only' double exponential length in
the size of the Petri net. This result leads to an algorithm which
requires at most space $2^{c n \log n}$ for some constant $c$. This
complexity is almost optimal, because Lipton proved \cite{Lipton76}
that deciding boundedness requires at least space $2^{c \sqrt{n}}$.

In \cite{RosierYen86}, Rosier and Yen carried out a multiparameter
analysis of the boundedness problem. They used three parameters:
$k$, the number of places; $l$, the maximum number of
inputs or outputs of a transition; and $n$, the number of transitions.
They refined Rackoff's result, and gave an algorithm that works in
$2^{ck\log k} (l + \log n)$ space. Among other results, they also showed
that, if $k$ is kept constant, then the problem is PSPACE-complete for $k
\geq 4$.

Boundedness can be decided at a lower cost for several classes of
nets. It is
\begin{itemize}
\item PSPACE-complete for single-path
Petri nets \cite{HowellJancarRosier93};
\item co-NP-complete for sinkless and normal Petri nets
\cite{HowellRosierYen93};
\item polynomial (quadratic) for
conflict-free Petri nets \cite{HowellRosierYen87}.
\end{itemize}

Some problems related to boundedness have also been studied. A Petri net
is  {\em $k$-bounded} if no reachable marking puts more than $k$ tokens
in any place (since we assume that the set of places of a net is finite,
$k$-bounded Petri nets are bounded). The $k$-boundedness problem is
PSPACE-complete \cite{JonesLandweberLien77}.

A net $N$ is {\em structurally bounded} if $(N, M)$ is bounded for all
possible markings $M$ of $N$. It can be shown that a net $N$ is
structurally bounded if and only if the system of linear inequations
$Y \cdot C \leq 0$, where $C$ is the so called
{\em incidence matrix} of $N$, has a solution \cite{MemmiRoucairol79}.
This result implies that the structural boundedness problem can be solved
in polynomial time using Linear Programming.

\paragraph{Reachability}

The reachability problem for Petri nets consists of deciding, given
a Petri net $(N, M_0)$ and a marking $M$ of $N$, if $M$ can be reached
from
$M_0$. It was soon observed by Hack \cite{Hack76} and Keller \cite{Keller72}
that many other problems were recursively equivalent to the reachability
problem, and so it became a central issue of net theory.
In spite of important efforts,
the problem remained elusive. Sacerdote and Tenney claimed in
\cite{SacerdoteTenney77} that reachability was decidable,
but did not give a complete
proof. This was not done until 1981 by Mayr
\cite{Mayr81}; later on, Kosaraju simplified the proof \cite{Kosaraju82},
basing on the ideas of \cite{SacerdoteTenney77} and \cite{Mayr81}.
The proof is very complicated. A detailed and self-contained
description can be found in Reutenauer's book \cite{Reutenauer90}, which
is devoted to
it. In \cite{Lambert92}, Lambert has simplified the proof further.

Petri nets can be added {\em inhibitor arcs} which make the firability
of a transition dependent upon the condition that a place contains no
tokens. It is well known that the reachability problem for Petri nets with
at least two inhibitor arcs is undecidable \cite{Hack76}. Very recently,
Reinhardt has shown that the problem for nets with only one inhibitor
arc is decidable \cite{Reinhardt94}.

Hack showed in \cite{Hack76} that several variations and
subproblems of the reachability problem are in fact recursively equivalent
to it:

\begin{itemize}
\item The submarking reachability problem. A submarking is a partially
specified markings (only the number of tokens that some of the places
 have to
contain is given). It can also be seen as the set of markings that
coincide on a certain subset of places. The problem consists of
deciding if some marking of this set is reachable.
\item The zero reachability problem. To decide if the zero marking
 -- the
one that puts no tokens in any place -- is reachable.
\item The single-place zero
reachability problem. To decide, given a place $s$, if there exists a
reachable marking which does not put any token on $s$.
\end{itemize}

The complexity of the reachability problem has been open for many
years. Lipton proved an exponential
space lower bound \cite{Lipton76}, while the known algorithms require
non-primitive recursive space. The situation is therefore similar
 to
that of the boundedness problem before Rackoff's result.
However, tight complexity bounds of the reachability problem are
known for
many net classes. Reachability is
\begin{itemize}
\item EXPSPACE-complete for
symmetric Petri nets; this result was first announced in
\cite{CardozaLiptonMeyer76}, and a proof was first given in
\cite{MayrMeyer82};
\item solvable in double exponential time for Petri nets with at most
five places \cite{HowellRosierHuynhYen86};
\item PSPACE-complete for nets
in which every transition has the same number of input and output places
\cite{JonesLandweberLien77};   \item PSPACE-complete for 1-safe Petri
nets \cite{ChengEsparzaPalsberg93};
\item PSPACE-complete for single-path
Petri nets \cite{HowellJancarRosier93};
\item NP-complete for Petri
nets without cycles \cite{Stewart92};
\item NP-complete for sinkless and normal Petri nets \cite{HowellRosierYen93};
\item NP-complete for conflict-free Petri nets \cite{HowellRosier88};
\item NP-complete for BPP-nets \cite{Huynh83,Esparza94b};
\item polynomial for bounded conflict-free Petri nets \cite{HowellRosier88};
\item polynomial for marked graphs
\cite{CommonerHoltEvenPnueli71,DeselEsparza94}; a Petri net is a {\em
marked graph} if every place has exactly one input transition and one
output transition (notice that marked graphs are conflict-free);
\item polynomial for live, bounded and cyclic free-choice nets
\cite{DeselEsparza93} (liveness is defined in the next paragraph).

\end{itemize}

\paragraph{Liveness}

Hack showed in \cite{Hack76} that the liveness problem is
recursively equivalent to the reachability problem (see also
\cite{ArakiKasami77a}), and thus decidable. Loosely
speaking, a Petri net is {\em live} if every transition can always
 occur
again; more precisely, if for every reachable marking $M$ and every
 transition
$t$, there exists an occurrence sequence $M \goes{\sigma} M'$ such
 that $M'$
enables $t$. The computational complexity of the liveness problem
 is open, but
there exist partial solutions for different classes. The liveness
 problem is
\begin{itemize}
\item PSPACE-complete for 1-safe Petri nets \cite{ChengEsparzaPalsberg93};

\item co-NP-complete for free-choice nets
\cite{JonesLandweberLien77};
\item polynomial for bounded free-choice nets \cite{EsparzaSilva92};
\item polynomial for conflict-free Petri nets \cite{HowellRosier89}.
\end{itemize}

\paragraph{Deadlock-freedom}

A Petri net is {\em deadlock-free} if
every reachable marking enables some transition. Deadlock-freedom
 can be
easily reduced in polynomial time to the reachability problem
\cite{ChengEsparzaPalsberg93}. The deadlock-freedom problem is:
\begin{itemize}
\item PSPACE-complete for 1-safe Petri nets, even if they are single-path
\cite{ChengEsparzaPalsberg93};
\item NP-complete for 1-safe free-choice Petri nets
\cite{ChengEsparzaPalsberg93};
\item polynomial for conflict-free Petri nets
\cite{HowellRosierYen87}.
\end{itemize}

\paragraph{Home states and home spaces}

A marking of a Petri net is a
{\em home state} if it is reachable from every reachable state. The
 home
state problem consists in deciding, given a Petri net $(N, M_0)$
 and a
reachable marking $M$, if $M$ is a home state. It was shown to be
decidable by Frutos \cite{Frutos86}. The subproblem
of deciding if the {\em initial} marking of a Petri net is a home
 state,
which is the problem of deciding if a Petri net is cyclic, was solved much
 earlier by
Araki and Kasami \cite{ArakiKasami77b}. The home state problem
is polynomial for live and bounded free-choice Petri nets
\cite{BestDeselEsparza92,DeselEsparza93}.

The home state problem is a special case of the home space problem. A set
of markings ${\cal M}$ of a Petri net is a {\em home space} if for every
reachable marking $M$, some marking of ${\cal M}$ is reachable from $M$.
The home space problem for linear sets is decidable
\cite{FrutosJohnen89} (for the definition of linear set, see the
semilinearity problem).

\paragraph{Promptness and strong promptness}

In a Petri net model of a system, the transitions represent the
atomic actions that the system can execute. Some of these actions
 may
be silent, i.e., not observable. A Petri net whose transitions
are partitioned into silent and observable is {\em prompt} if every
infinite occurrence sequence contains infinitely many observable
transitions. It is {\em strongly prompt} if there exists a number
 $k$ such
that no occurrence sequence contains more than $k$ consecutive silent
transitions. Promptness is thus
strongly related to the notion of divergence in process algebras.
The promptness and strong promptness problems were shown to be
decidable by Valk and Jantzen \cite{ValkJantzen85}. It follows easily
 from
a result of \cite{ThiagarajanVoss84} that the promptness problem
 is polynomial
for live and bounded free-choice Petri nets.

\paragraph{Persistence}

The persistence problem (to decide if a given Petri net is persistent)
was shown to be decidable by Grabowsky \cite{Grabowsky80}, Mayr
\cite{Mayr81} and M\"{u}ller \cite{Muller81}.
 It is
not known if the problem is primitive recursive. It is PSPACE-complete
 for
1-safe nets \cite{ChengEsparzaPalsberg93}.

\paragraph{Regularity and context-freeness}

The regularity and context-freeness problems are in fact a
collection of problems of the form:
\begin{quote}
to decide if the X of a Y-Petri net is Z
\end{quote}
where X can be `trace set' or `language', Y
can be `labelled' or `unlabelled', and Z can be `regular'
or `context-free'. Ginzburg and Yoeli
\cite{GinzburgYoeli80} and Valk and Vidal-Naquet
\cite{ValkVidal-Naquet81} proved independently that the regularity
problem for
trace sets of unlabelled Petri nets is decidable (see also
\cite{Schwer86}). Other results of \cite{ValkVidal-Naquet81}
are that this problem is not primitive recursive,  and that the
regularity problem for languages of labelled Petri nets is undecidable
(see also \cite{Jantzen79}).

The decidability of the context-freeness problem for trace
sets of unlabelled Petri nets has been proved by Schwer \cite{Schwer92}.

\paragraph{Semilinearity}

Markings can be seen, once an
arbitrary ordering of the set of places is taken, as vectors over
$\nsymbol^n$, where $n$ is the number of places of the net. A subset
 of
$\nsymbol^n$ is {\em linear} if it is of the form $$\{ u + \sum_{i= 1}^p
 n_i
v_i \mid n_i \in \nsymbol \}$$
\noindent where $u, v_1, \ldots, v_p$ belong to $\nsymbol^n$. A subset
 of
$\nsymbol^n$ is {\em semilinear} if it is a finite union of linear
 sets.

Some interesting problems are decidable
for Petri nets whose set of reachable markings is semilinear. Many net
subclasses, unfortunately all of them quite restrictive, are known to
have semilnear reachability sets, as we shall see in section
\ref{equivalences}. It was also proved by Kleine B{\"u}ning, Lettmann and
Mayr that the {\em projection} of the set of
reachable markings on a place of the net is always
semilinear \cite{KleineLettmannMayr89}.

The semilinearity problem is the problem of deciding if the set of
reachable markings of a given Petri net is semilinear. Its
decidability was proved independently by Hauschildt
\cite{Hauschildt90} and Lambert \cite{Lambert94}.

\paragraph{Non-termination}

Much effort has been devoted to the decidability of termination
in Petri nets under fairness conditions. This study was
initiated by Carstensen \cite{Carstensen87}, where he proved that
 the
fair non-termination problem is undecidable. An infinite occurrence
sequence is {\em fair} if a transition which is enabled at infinitely
many markings of the sequence appears infinitely often in it. If every
maximal fair occurrence sequence of a Petri net is finite (i.e., it ends
at a deadlocked marking), then we say that the Petri net is guaranteed to
{\em terminate} under the fairness assumption. The
fair non-termination problem consists in deciding if a given Petri net
is {\em not} guaranteed to terminate, i.e., if it has an
infinite fair occurrence sequence.

In \cite{HowellRosierYen91}, Howell, Rosier and Yen conducted an
exhaustive study of the decidability and complexity of
non-termination problems for 24 different fairness notions. In
particular, they studied the three notions of impartiality,
justice and fairness introduced in \cite{LehmanPnueliStavi81}. An
infinite occurrence sequence is {\em impartial} if every transition
 of
the net occurs infinitely often in it; it is {\em just} if every
transition that is enabled almost everywhere along the sequence occurs
infinitely often in it; fair infinite occurrence sequences were
defined above. The just
non-termination problem was left open in \cite{HowellRosierYen91},
 and was
later solved by Jan\v{c}ar \cite{Jancar90}. The final picture is
 the
following:

\begin{itemize}
\item The fair non-termination problem is complete for the first
level of the analytical hierarchy. The restriction of this problem
 to
bounded Petri nets is decidable,
but non-primitive recursive.
\item The impartial non-termination problem can be reduced in
polynomial time to the boundedness problem, and can therefore be
 solved in
exponential space.
\item The just non-termination problem is decidable, and at
least as hard as the reachability problem.
\end{itemize}

Two other interesting results of \cite{HowellRosierYen91} concern
 the notions
of $i$-fairness and $\infty$-fairness introduced by Best \cite{Best84}.
 A
transition $t$ is {\em $i$-enabled} at a marking if there is an occurrence
sequence no longer than $i$ transitions which enables $t$. A
transition is {\em $\infty$-enabled} at a marking if there is an
 occurrence
sequence, no matter how long, which enables $t$. An
infinite occurrence sequence is {\em $i$-fair} ({\em $\infty$-fair})
 if every
transition which is infinitely often $i$-enabled
($\infty$-enabled) along the sequence occurs infinitely often in
 it.

Observe that $0$-fairness coincides with fairness in the sense of
\cite{Carstensen87} and \cite{HowellRosierYen91}. Therefore, the
 $0$-fair
non-termination problem is undecidable. The $i$-fair non-termination
 problem
turns out to be undecidable as well for every $i$. However, the
$\infty$-fair non-termination problem is reducible in polynomial time to
the reachability problem, and thus decidable.

\section{Equivalences}
\label{equivalences}
As opposed to the results from the previous section, the main message
from the study of
decidability of behavioural equivalences of Petri nets is that almost
 all
results are negative.
However, many interesting and nontrivial subclasses of nets have
been identified for which
equivalences become decidable, thus shedding some light on the sources
of the complexity of
net behaviours.

The first undecidability result for equivalences of Petri nets dates
 back
to the early
seventies.
\paragraph{Marking equivalence}
Two Petri nets having the same set of places are
said to be {\em marking equivalent} iff they have the same set of
 reachable
markings.

Marking equivalence is undecidable for Petri nets.
This result was proved by Hack \cite{Hack76}, using a former result
 by
Rabin, proving that the marking
inclusion problem is undecidable. (Rabin never published his result;
for a description of the proof, see \cite{Baker73}). The idea relies on a
rather subtle way of computing functions by nets in a weak sense. It is
 then proved
that diophantine
polynomials may be computed, and then Hilbert's tenth
problem is reduced to the marking inclusion/equivalence problem.

The more straightforward approach to prove undecidability, by attempting
 to
simulate some
universal computing device like Counter Machines by nets (representing
counters and
their values by places and their number of tokens) fails because
 of the
inability of nets to
``test for zero''. But there is an obvious and simple way of
semi-simulating Counter
Machines by nets, simulating the counter-manipulations step by step,
 {\em but}
allowing some computational branches conditioned on a counter having
 the value zero
to be followed in the simulation, even though the corresponding place
 is
nonempty.

 Recently,
Jan\v{c}ar \cite{Jancar93} came up with a set of ingenious, simple
 and
elegant proofs of
undecidability of equivalence problems following the pattern:
\begin{quote}
to prove undecidability of $X$-equivalence, construct two modifications
of the simple
nets semi-simulating a given Counter Machine, CM, satisfying that
CM halts iff the two
constructed nets are not $X$-equivalent.
\end{quote}
(actually, the first proof of this kind can be found, to our knowledge,
 in
\cite{ArakiKasami77a}, but Jan\v{c}ar has generalized the principle
 to other
equivalences). In \cite{Jancar93} the reader may find a simple and elegant
 proof of
undecidability of marking equivalence (among others) for nets following
exactly this pattern. It shows that the problem is undecidable even
 for nets
with five unbounded places (i.e., places $s$ such that for every number
 $k$ there
exists a reachable marking $M$ such that $M(s) > k$).

For certain restricted classes of nets the marking equivalence
 problem
has been
shown to be
decidable. For instance, it was noticed very early that for nets
with a semilinear set of reachable
markings the problem is decidable. This is due to a connection between
semilinear sets and
Pressburger arithmetic, a decidable first order theory.
And many nontrivial restricted classes of Petri nets have been shown
to have effectively
computable semilinear reachable markings. A few examples:
\begin{itemize}
\item persistent
\cite{LandweberRobertson75,Grabowsky80,Mayr81,Muller81} and weakly
persistent nets
\cite{Yamasaki81};
\item nets with at most five places
\cite{HopcroftPansiot78} (there exist nets with six places having
 a non-semilinear reachability set);
\item regular nets
\cite{GinzburgYoeli80,ValkVidal-Naquet81}; a Petri net is {\em regular}
 if its
trace set is regular;
\item cyclic nets \cite{ArakiKasami77a};
\item BPP-nets \cite{Esparza94b};
\end{itemize}

For some classes, the complexity of the problem has been determined.
 It is:
\begin{itemize}
\item solvable in $c^{n \log n}$ space for symmetric Petri nets
\cite{Huynh85};
\item solvable in double exponential time for nets with at most five
places \cite{HowellRosierHuynhYen86};
\item $\Pi_2^P$-complete for
conflict-free Petri nets, where $\Pi_2^P$ is the class of languages whose
complements are in the second level
of the polynomial-time hierarchy \cite{HowellRosier88};
\item $\Pi_2^P$-complete for sinkless and normal Petri nets
\cite{HowellRosierYen93};
\item PSPACE-complete for single-path Petri nets
\cite{HowellJancarRosier93}.
\end{itemize}

Also, the marking equivalence problem is obviously decidable for
bounded nets, which only have finitely many reachable markings.
It was shown by Mayr and Meyer \cite{MayrMeyer81} that the problem
is not primitive recursively decidable.
This result has since been strengthened by McAloon \cite{McAloon84}
 and Clote
\cite{Clote86},
who showed that it is complete for
DTIME in the Ackerman-\-function.
McAloon also showed that the restriction of the problem to Petri nets
with at most a fixed number $k$ of places is primitive recursive. The
restriction to 1-safe Petri nets is PSPACE-complete
\cite{ChengEsparzaPalsberg93}.

Most - if not all - of these results also hold for the inclusion problem.

\paragraph{Trace and language equivalences}
Another bulk of results are concerned with equivalences of nets in
terms of occurrence sequences.
Two (labelled) Petri nets are said to be {\em trace equivalent} ({\em
 language
equivalent}) if they have the same trace set (language).
Hack proved in \cite{Hack76} that the problems of deciding if two
 labelled Petri
nets are language equivalent or trace equivalent are undecidable,
 by means
of a reduction from
the marking equivalence problem.
Araki and Kasami gave another proof \cite{ArakiKasami77a}
by reduction from the halting problem for Counter Machines. Stronger
results are:
\begin{itemize}
\item trace equivalence is undecidable for labelled Petri nets with
 at most two
unbounded places \cite{Jancar93};
\item language equivalence is undecidable for labelled Petri nets,
 one of them
having one unbounded place and the other none \cite{ValkVidal-Naquet81};
\item trace and language equivalence are undecidable for
BPP-nets \cite{Hirshfeld93}.
This is a remarkable result, since BPP-nets are a class with  rather
 limited
expressive power.
\end{itemize}
The trace equivalence problem of Petri nets with
exactly one unbounded place is, to the best of our knowledge, open.

If we restrict ourselves to unlabelled nets,
both problems become decidable.
Hack \cite{Hack76} gave a reduction to the reachability problem,
 and hence
today we
conclude decidability.

It is well-known that any trace set of a labelled net is also a language
of some labelled net,
but not vice versa.
This raises the interesting question, whether there exists some class
of nets which
distinguishes the two equivalence problems with respect to decidability.

A labelled net is said to be {\em deterministic up to bisimilarity}
 iff for
all reachable
markings $M$, if two transitions $t'$ and $t''$ carrying the same
 label are
enabled, $M\goes{t'} M'$
and $M\goes{t''}M''$, then $M'$ and $M''$ are strongly bisimilar
(for the definition of strong bisimulation, see below).

Clearly any unlabelled net is deterministic up to bisimilarity, but
not vice
versa. Further\-more, it has been shown that the property of being
deterministic up to
bisimilarity is decidable (reduced to the reachability problem in
\cite{Jancar93}).
Christensen has shown \cite{Christensen93} that for nets which
are deterministic up to
bisimilarity, trace equivalence is indeed decidable, but language
equivalence is not!

\paragraph{Bisimulation equivalence}
This brings us to the question of {\em bisimulation equivalence}
\cite{Milner89} for nets. We recall the definition of bisimilar
markings and bisimilar nets. A relation
$R$ between the nodes of two labelled graphs
is a {\em (strong) bisimulation} if it is symmetric, and for every element
$(n_1, n_2)$ of $R$, the following condition holds:
\begin{quote}
if $n_1 \goes{a}
n_1'$, then there exists a node $n_2'$ such that $n_2 \goes{a}
n_2'$ and $(n_1', n_2')$ belongs to $R$.
\end{quote}
Let $(N_1, M_{01})$ and $(N_2, M_{02})$ be two Petri nets, let
${\cal R}_1$, ${\cal R}_2$ be their reachabilty graphs, and let $M_1$,
$M_2$ be nodes of ${\cal R}_1$ and ${\cal R}_2$, respectively.  We say
that $M_1$ and $M_2$ are (strongly) bisimilar if some bisimulation between
the nodes of ${\cal R}_1$ and ${\cal R}_2$ contains the pair
$(M_1, M_2)$. The Petri nets $(N_1, M_{01})$ and $(N_2, M_{02})$ are
(strongly) bisimilar if the initial markings $M_{01}$ and $M_{02}$ are
bisimilar.

Notice that the
definition of the reachability graph is different for labelled and
unlabelled nets, and therefore the corresponding notions of bisimulation
also differ. It is easy to see that for unlabelled nets bisimulation
and trace equivalance coincide. For labelled nets, bisimulation
equivalence implies trace equivalence, but not vice versa.

Some results for this
problem are:
\begin{itemize}
\item undecidable for labelled nets, even with only two
unbounded places \cite{Jancar93}, proof following the
``Jan\v{c}ar-pattern''
\cite{Jancar93};
\item decidable for labelled BPP-nets \cite{ChristensenHirshfeldMoller93};
\item decidable for labelled nets, if just one of them is deterministic
 up to bisimulation
 \cite{Jancar93};
\item decidable for unlabelled nets (because trace equivalence is
decidable, and bisimulation and trace equivalence coincide).
\end{itemize}

\paragraph{Other equivalences}
H\"{u}ttel has recently shown in \cite{Huttel93} that all the
equivalences of the linear/branching time hierarchy \cite{vanGlabbeek90}
 below
bisimulation equivalence are undecidable for Basic Parallel Processes.
 This
result implies that they are undecidable for labelled BPP-nets. Together
with the undecidability of bisimulation for labelled Petri nets,
 we then have
that all the interleaving equivalences described so far in the literature
 are
undecidable.

On the other hand, all problems from the linear/branching time hierarchy
become decidable if we
restrict ourselves to bounded nets. The complexity of these problems
 has been
studied by
several people, and some of the clever algorithms invented are parts
 of
various
constructed tools for reasoning about concurrent computations. Here
 we just
mention
the following results from \cite{JategaonkarMeyer93} for 1-safe
nets:
\begin{itemize}
\item the language and trace equivalences are both complete for EXPSPACE;
interestingly, the same complexity result holds for their ``true
 concurrency''
counterparts in terms of (Pratt-)pomset equivalences;
 \item the bisimulation  equivalence is complete for DEXPTIME;
interestingly, the
same complexity result holds for its ``true concurrency'' counterparts,
like {\em
history preserving bisimulation} \cite{RabinowichTrakhtenbrot88}.
 \end{itemize}

\section{Temporal Logics}
\label{temporallogics}
The very positive balance of section \ref{properties} (in spite of
 the
considerable expressive power of Petri nets, most properties are
 decidable),
has encouraged researchers to study decidability issues for specification
languages in which a large set of properties can be expressed.  Mostly,
 these
languages take the shape of a {\em temporal logic}. The problem of
deciding, given a Petri net and a formula of a temporal logic, if
 the
net satisfies the formula, is called the {\em model checking}
problem.

Temporal logics can be classified into two groups: {\em linear time}
and {\em branching time} logics. Linear time logics for Petri nets
are usually interpreted on the set of maximal occurrence
sequences\footnote{Other equivalent interpretations are also used.}.
Branching time
logics are interpreted on the reachability graph. It is well
known that some properties can be more naturally expressed in a linear
time logic than in a branching time one, and vice versa.

The results on
branching time temporal logics are mostly negative. Esparza shows in
\cite{Esparza94a} that the model checking problem for (a Petri net
version of) the logic UB$^-$ \cite{EmersonSrinivasan89} is undecidable.
This is one of the weakest branching time logics described in the
literature. It has basic predicates of the form $ge(s, c)$, where $s$ is
a place of the net and $c$ is a nonnegative constant. A predicate
$ge(s,c)$ is read `the number of tokens of $s$ is greater than or equal
to $c$'; accordingly, it holds at a marking $M$ if $M(s) \geq c$. The
operators of the logic are the usual boolean connectives, ${\bf EX}$
(`existential next') and ${\bf EF}$ (`possibly'). A reachable marking
satisfies a property ${\bf EX}\phi$ if it enables some transition $t$
and the marking reached by the occurrence of $t$ satisfies $\phi$; a
marking satisfies ${\bf EF}\phi$ if it enables an occurrence
sequence $\sigma$ such that some marking visited along the execution
of
$\sigma$ satisfies $\phi$ \footnote{The logic described in
\cite{Esparza94a} is in fact slightly weaker than UB$^-$. We have chosen
it to better compare results.}.

UB$^-$ is decidable for any net whose
set of reachable markings is effectively semilinear, because the model
checking problem can be then reduced to the satisfiability problem of
the first order logic of the natural numbers with addition, also known
as Presburger Arithmetic. This includes, for instance, BPP-nets
or conflict-free nets. For 1-safe conflict-free nets it is even
decidable in polynomial time \cite{Esparza93} (for the subclass of
1-safe marked graphs the same result had been proven in
\cite{BestEsparza92}).

The logic UB is obtained by adding the operator ${\bf EG}$ to UB$^-$. A
marking satisfies a property ${\bf EG} \phi$ if it enables some
infinite occurrence sequence $\sigma$ and {\em all} the
markings visited along the execution of $\sigma$ satisfy $\phi$. Esparza
has recently showed that UB is undecidable for BPP-nets
\cite{Esparza94c}. The result can be trasferred to Basic Parallel
Processes.

Other branching temporal logics, such as CTL and CTL$^\star$
\cite{EmersonSrinivasan89}, or the $mu$-calculus \cite{Mu-calculus}, are
more expressive than UB, and therefore the undecidability results carry
over (see also \cite{Bradfield91}).

The conclusion that can be derived is that no
natural and useful branching time temporal logic for Petri nets seems
to be decidable.

There has been more research on linear time temporal logics for Petri
nets. To provide a unifying
framework in which to survey the results we add two more basic
predicates to the predicates
$ge(s, c)$, and then build different temporal logics on top of them. The
predicates are now interpreted on the markings of a maximal occurrence
sequence. We say that an occurrence sequence satisfies a formula of a
logic if its initial marking satisfies it. Finally, a Petri net satisfies
a formula if at least one maximal occurrence sequences satisfy
it (or, equivalently, if every maximal occurrence sequence satisfies its
negation). The new predicates are:
\begin{itemize}
\item ${\em first}(t)$, where $t$ is a transition of the net. It
 holds
at a marking $M$ if the transition that succeeds $M$ in the occurrence
sequence is $t$.
\item $en(t)$, where $t$ is a transition of the net. It holds
at a marking $M$ if $M$ enables $t$\footnote{The predicate $en(t)$
can be derived as the conjunction of $ge(s, 1)$ for every input place
 $s$
of $t$. We include it as a basic predicate for convenience.}.
\end{itemize}

Esparza shows in \cite{Esparza94a} that the linear time $\mu$-calculus
\cite{linear-Mu-calculus} with ${\em first}(t)$ as only basic
predicate is decidable.   If the predicates $ge(s, c)$ are added,
then the logic becomes undecidable, even for BPP-nets.

Howell and Rosier studied in
\cite{HowellRosier89} the logic with all three basic predicates and
 an eventuality operator ${\bf F}$, where a marking of an occurrence
 sequence
satisfies ${\bf F} \phi$ if some later marking satisfies $\phi$.
They showed that
the model checking problem is undecidable, even for conflict-free
 Petri
nets (notice that the fair
non-termination problem can be reduced to the model checking
problem for this logic: a Petri net satisfies
the formula ${\bf G}{\bf F} en(t) \Rightarrow {\bf G}{\bf F} {\it
first}(t)$, where ${\bf G} =  \neg {\bf F} \neg$, if some occurrence
sequence that enables $t$ infinitely often contains $t$ infinitely often).
It follows from results of \cite{Esparza94c} that it is also
undecidable for BPP-nets.

The model checking problem is, however, decidable for two
fragments:
\begin{itemize}
\item The fragment in which negations are only applied to
predicates \cite{HowellRosierYen91}.

This fragment contains the formula ${\bf
F}{\it first}(t)$, which expresses that $t$ eventually
occurs, but not ${\bf
GF}{\it first}(t)$, which expresses that $t$ is bound to occur
infinitely often. The model checking problem for this fragment can
 be
reduced in polynomial time to the reachability problem. For the class
of conflict-free nets, the model checking problem is NP-complete.

\item The fragment in which the composed operator ${\bf G}{\bf
F}$ is the only one allowed, and negations are only applied
to predicates \cite{Jancar90}.

This fragment contains the formula ${\bf G}{\bf F} {\it
first}(t)$, but not, for instance, the formula ${\bf G}{\bf F} {\it
first}(t) \Rightarrow {\bf G}{\bf F} {\it first}(t')$ (after replacing
 the
implication by its definition, a negation appears in front of an
 operator).
Jan\v{c}ar \cite{Jancar90}  reduces the model checking problem
 for this
fragment to an exponential number of instances
of the reachability problem. If the formula is of the form ${\bf
 G}{\bf
F} \phi$, where $\phi$ is a boolean combination of basic predicates,
then a better result exists: the model checking problem can be reduced
in polynomial time to the reachability problem \cite{HowellRosierYen91}.
\end{itemize}

These results show that the presence or absence of place predicates
 is
decisive for the decidability of a linear time logic. When they are
 absent,
even rather powerful logics as the linear time $\mu$-calculus are
 decidable.
When they are present, no natural logic is decidable, only fragments
 in which
some restrictions are applied to the use of boolean connectives.

All the decidable fragments of these logics are at least as hard
 as the
reachability problem, which, as mentioned in the first section, is
EXPSPACE-hard, and could well be non-primitive recursive. Yen has
 defined
in \cite{Yen92} a class of path formulas which can be decided in
exponential
 space.
The class is of the form
$$
\begin{array}{ll}
\exists \: M_1, M_2, \ldots, M_k \; \exists \: \sigma_1, \sigma_2,
\ldots, \sigma_k & (M_0 \goes{\sigma_1} M_1 \goes{\sigma_2} M_2
 \ldots
\goes{\sigma_k} M_k) \\
&  \wedge \: F(M_1, \ldots, M_k, \sigma_1, \ldots,
\sigma_k) \end{array}
$$
\noindent where $F$ belongs to a certain set of predicates. This
 set includes
arbitrary conjunctions and disjunctions of both place predicates,
 such as
\begin{itemize} \item $M(s) \geq c$ for a marking $M$,
place $s$ and constant $c$,
\item $M(s) \geq M'(s) + c$, for markings $M$ and $M'$, place $s$
 and
constant $c$,
\end{itemize}
and transition predicates, such as
\begin{itemize}
\item $\#_\sigma (t) \geq c$ for a
transition sequence $\sigma$, transition $t$ and constant $c$, which
 is true
if the sequence $\sigma$ contains $t$ at least $c$ times.
\end{itemize}

Recently, Yen, Wang and Yang have shown that deciding this class
 of formulas is
NP-complete for sinkless nets and polynomial for conflict-free nets
\cite{YenWangYang93}.

\section*{Acknowledgements}
Thanks to David de Frutos, Matthias Jantzen, Jean-Luc Lambert,
Elisabeth Pelz, Hsu-Chun Yen, and two anonymous referees for providing
very useful information.

\end{document}